\documentclass[12pt]{article}
\usepackage{graphicx}
\usepackage{cite}
\usepackage{amsmath}
\usepackage{amssymb}
\usepackage{enumerate} 
\usepackage{latexsym}
\usepackage{color}
\usepackage{cite}
\usepackage{graphicx}
\usepackage{float}
\usepackage[caption = false]{subfig}
\begin{document}
\vskip 2cm

\def\warning#1{\begin{center}
\framebox{\parbox{0.8\columnwidth}{\large\bf #1}}
\end{center}}

\begin{center}
		{\bf\Large{ Analysis of the Fokker-Planck Equation in Schwarzschild Spacetime: A Supersymmetric Connection} }
		\vskip 1.5 cm
{\sf{ \bf  Ojaswini Sharma and  Aradhya Shukla$^a$}}\\
\vskip .1cm
{\it $^a$Department of Physics, Institute of Applied Sciences and Humanities, \\GLA 
University, Mathura- 281406, Uttar Pradesh , India}\\
\vskip .25cm
		{E-mails: {\tt ojaswini.sharma\_phd24@gla.ac.in, ashukla038@gmail.com}}
		
		\vskip 1cm
\end{center}
\vskip 1cm
	\noindent
	\textbf{Abstract:} We have re-analyzed the dynamics of the thermal potential within Schwarzschild spacetime by employing the Fokker-Planck equation. We demonstrate that the Fokker-Planck equation reduces to a simplified form equivalent to a scaled quantum mechanical problem with a harmonic oscillator potential. In this framework, we highlight an interesting correspondence between supersymmetric quantum mechanics (SUSY QM) and the Fokker-Planck dynamics associated with the Schwarzschild metric. Utilizing the isospectral deformation, an intrinsic feature of SUSY QM, we derive a family of one-parameter isospectral potentials. Notably, this new class of potentials exhibits the same energy spectrum as the original harmonic oscillator potential, but with distinct wavefunctions.
    
\date{}

\section{Introduction}
The supersymmetry (SUSY) has first been conceptualized in the context of field theory \cite{wb}. There are two families of particles: fermions and bosons. Fermions are basic building blocks of matter; whereas, bosons mediate the interactions \cite{wei1}. The SUSY theory interrelates fermion with boson i.e. for each boson there will be a fermion and vice versa \cite{pol, wei2}. However, SUSY has not yet been experimentally verified \cite{man}. Drawing inspiration from field-theoretic SUSY, a quantum-mechanical version of SUSY has been developed \cite{wit,cooper1}. In SUSY quantum mechanics (QM), there are two different potentials known as supersymmetric partners, which share identical energy spectra \cite{khare1, dutt}. SUSY is said to be unbroken when both partner potentials have a common ground state \cite{cooper2}, and it is considered broken if the two potentials possess different ground states \cite{asim}.

The SUSY QM has the unique property of interrelating two different Hamiltonians. In this formalism, Hamiltonian is factorized into a pair of SUSY partner Hamiltonian \cite{khare1}. As a result, even a trivially solvable potential can give rise to a class of non-trivial potentials, and vice versa. For instance, a free particle has a partner potential, which can be a periodic Scarf potential or a finite-depth Pöschl-Teller potential \cite{scarf, pos, ara1}. In addition, SUSY QM offers many intriguing features, such as shape invariance \cite{asim1, gen, bou} and isospectral deformation \cite{cooper1}. Shape invariance, where the partner potentials maintain the same functional form, enables the determination of a complete spectrum without solving multiple equations \cite{asim1}. On the other hand, isospectral deformation generates a family of potentials with identical energy spectra as the original potential \cite{mie}. Moreover, isospectral deformation proves valuable in reconstructing unknown potentials, particularly in inverse scattering problems and imaging \cite{nov, suku}. The process of deformation of the potential can be done by two different methods; translation \cite{asim} and scaling of the superpotential \cite{jen}. Recently, two parameter and three parameter isospectral potentials have been constructed for reflectionless potential by scaling the methodology \cite{ara2}. However, it has been shown that both the methods physically yield similar result \cite{ara2, roy}.

The SUSY QM are extremely useful in studying complex quantum mechanical and optical systems \cite{kla, mac}. The connection between SUSY QM and the Helmholtz equation in optics is well-established \cite{kuz}. By employing isospectral deformation within SUSY, one can generate a family of refractive index profiles for optical systems and photonic crystals that possess identical optical properties, such as reflection and refraction \cite{miri, garc}.

The concepts of SUSY QM have been extended to curved spacetime and cosmology \cite{mon, pau}. Recently, the shape invariance property has been utilized in the spatially flat Friedmann-Robertson-Walker (FRW) model \cite{jrm1}, playing a crucial role in the formulation of the associated Wheeler-DeWitt equation \cite{jrm2}. Furthermore, a novel connection has been established between stochastic processes in curved configuration space and SUSY QM \cite{GR1}. The integration of SUSY Grassmann variables into stochastic processes has also proven to be valuable in determining the physical Lyapunov exponents \cite{GR2}, providing deeper insights into the perturbations in spacetime and the motion of particles within these curved geometries. Moreover, the stochastic evolution of the probability distribution is governed by the Fokker-Planck (FP) equation, which has been extended to model diffusion phenomena in curved spacetime \cite{malda, CMB, LHX}. The FP equation provides important insights into the gravitational and thermal effects on the random motion of particles near black holes \cite{zw, ris}. Recently, it has been demonstrated that within the FP framework \cite{xu}, the thermal potential of a Schwarzschild black hole is equivalent to a quantum mechanical oscillator, with discrete energy levels proportional to the temperature of the ensemble. By applying the formalism of SUSY quantum mechanics, we derived a new potential that has the same energy spectra but different wavefunctions \cite{fred}. This finding is particularly interesting, as the probability density for the particle’s motion in this quantum system differs significantly from the previous one.

 In Section 2, we provide a brief discussion on the FP equation for the Schwarzschild metric. The basic formalism of SUSY QM and isospectral deformation is outlined in Section 3 for completeness. Section 4 is dedicated to establishing the connection between SUSY QM and the FP equation, and generating a new one parameter isospectral potential. Finally, we conclude in Section 5.

\section{Schwarzschild Spacetime and Fokker-Planck's Equation}
The phase transition in black holes driven by thermal fluctuations can be modeled as a stochastic process. The evolution of the probability distribution for the black hole states is effectively analyzed using the FP formalism. The FP equation, assuming a constant diffusion coefficient 
$D$ and a time-independent drift coefficient $\mu(x)$, is expressed as \cite{ris}
\begin{eqnarray}
\frac{\partial P(x,t)}{\partial t}&=&\Big[\frac{\partial}{\partial x}f'(x)+D\frac{\partial^2}{\partial x^2}\Big]P(x,t)\nonumber\\
&=& L_{FP}P(x,t)=-\frac{\partial S(x,t)}{\partial x},
\end{eqnarray}
here, $f(x)= -\int^x \mu(y)\,dy$ represents the potential term with $f'(x)= \frac{df(x)}{dx}$ and $S(x,t)$ denotes the probability current.
By assuming an ansatz for the probability density, $P(x,t)=\Phi(x)e^{E\, t}$ and applying the relevant boundary conditions, the FP eqn. is transformed into an eigenvalue equation
\begin{equation}
    L_{FP}\,\Phi(x)= -E\,\Phi(x).
\end{equation}
By scaling the potential function with the diffusion coefficient,  $\eta(x)=f(x)/D$, Eq. (2) transforms into the eigenvalue equation $L\Psi(x)= E\,\Psi(x)$
where, $\Psi(x)=e^{\eta(x)/2}\Phi(x)$ and  operator $L$ is then given by
\begin{equation}
    L=-D\frac{\partial^2}{\partial x^2}+ V (x),\quad V (x)=\frac{1}{4D}[f'(x)]^2-\frac{1}{2}f''(x),
\end{equation}
having the similar form of Hamiltonian operator in Schr\"odinger equation.

The metric for (3+1)-dimensional Schwarzschild black hole is given by \cite{LHX}
\begin{equation}
   \hskip -0.3cm ds^2=-\Big(1-\frac{2m}{r}\Big) dt^2+\frac{dr^2}{1-2m/r}+r^2(d\theta^2+sin^2\theta d\phi^2),
\end{equation}
in the above $m$ represents the ADM mass of the black hole. The Schwarzschild metric has a spherical symmetry with singularities at $r=0$ and $r = 2m$. Thermal potential  for this black hole can be calculated as: $  U=\frac{1}{2}r_h-\pi Tr^2_h$ with event horizon $(r_h)$ and  is the ensemble temperature ($T$). It is to be noted that the thermal potential reaches its maximum value at $T = T_h$.

For analyzing the dynamical characteristic of thermal potential for Schwarzschild black hole within the framework of FP eqn.,  the potential is considered as $f(x)= \frac{1}{2}x-\pi T x^2$, yielding the effective potential as\\
\begin{equation}
    V(x)=\Big(\pi^2T^2D\Big) y^2+\pi T, \quad y=x- \Big(1/4\pi T\Big).
\end{equation}
From eqns. (3) and (4), one can easily get  the  FP eqn. in the  simplest form as 
\begin{equation}
   \pi T \Big[-\frac{\partial^2}{\partial \zeta^2}+\big(1+\zeta^2\big) \Big]\Psi(x)= E\, \Psi(x),
\end{equation}
where, $\zeta= \sqrt{\pi T/D}\,y$ is an auxiliary variable.
The above equation is the quantum mechanical version of harmonic oscillator modulo a constant factor.  From the above equation, one can get the energy eigenvalue and eigenfunction as
\begin{eqnarray}
&&E_n=2\pi T(n+1)\nonumber,\qquad \\ 
 &&\Psi_n(x)=\Big(\frac{T}{D}\Big)^{1/4}\frac{1}{\sqrt{2^n\,n!}}\,H_n(\zeta)\,e^{-\zeta^2/2},\quad \\ \nonumber
&&\text{with} \quad n = 0, 1, 2...
\end{eqnarray}
in the above $H_n(\zeta)$ are the Hermite polynomials. 

Our goal in the next section is to make a connection between the Fokker-Planck’s dynamic of thermal potential for Schwarzschild metric and supersymmetric quantum mechanics. We briefly discuss about the basic property of SUSY QM and isospectral deformation in SUSY QM which generates a new class of potential corresponding to the original one.

\section{Isospectral Deformation in Supersymmetric Quantum Mechanics}
We start with a Hamiltonian for any given potential $V(x)$.
\begin{equation}
    H= - \frac{d^2}{dx^2}+V,
\end{equation}
one can generate two superpartner Hamiltonians connected with each other as following \cite{cooper1,dutt},
\begin{eqnarray}
    H_+=A^+A, \qquad H_-=AA^+,
\end{eqnarray}
where,
\begin{equation}
    A=\frac{d}{dx} + W(x), \qquad  A^{\dagger} =-\frac{d}{dx} + W(x),
\end{equation}
 with the superpotential $W(x)$ which is connected with the potential $V(x)$. The two different forms of $H$ i.e., $H_+$ and $H_-$ \cite{mie, ara1} can be found by putting the values of $A$ and $A^\dagger$ in eqn. (9).
 
 \begin{equation}
     H_+= -\frac{d^2}{dx^2}+ W^2+W',
 \end{equation}
\begin{equation}
    H_-=-\frac{d^2}{dx^2}+ W^2-W',
\end{equation}
in the above prime denotes the differentiation w.r.t. space coordinate. After comparing eqn. (8), (11) and (12) we get the partner potentials $V_+$ and $V_-$ in terms of superpotential $W$ given below \cite{asim,ara2}
\begin{equation}
    V_+= W^2+W',  \qquad V_-= W^2-W'.
\end{equation}
It is to be noted that $V_+$ and $V_-$ have the same energy eigenvalue with common ground state.

A continuous alteration of the criteria of a system in such a way that its spectrum remains the same is one of the unique methodologies to study the non-trivial QM systems. 
In SUSY QM, isospectral deformation is used to create a new potential with same eigenspectra as the original potential. One can generate a family of physically different but spectrally similar Hamiltonians . In the isospectral deformation, the superpotential corresponding to the given partner potential is translated by a general function of the space.  Remarkably, it is not necessary that the isospectral potential have to be the SUSY partner potential of the original potential.  We translate the original potential as \cite{asim},
 \begin{equation}
     W_1(x)= W(x)+f(x),
 \end{equation}
where, $f(x)$ is a space dependent function. Putting eqn. (14) in (13) we get a condition,
\begin{equation}
        f(x)^2+2W(x)f(x)+\frac{df}{dx}=0,
 \end{equation}
 known as Bernouli equation. The above equation has a solution of the form
\begin{equation}
        f(x) =\frac{e^{-\int 2W(x) dx}}{{\lambda_1 -  \int e^{-\int 2W(x) dx}dx}},
 \end{equation}
 with $\lambda$ as an integration constant. Under this condition $V_-$ remains intact and a new form of potential $V_+$ is generated. One should note that the isopectral deformation only acts on the boundaries of the systems leaving the original potential consistent. It allows one to develop a completely distinct family of potential which may not be SUSY equivalent of original potential  \cite{dutt}.

 In the following sections, we discuss about the dynamics of FP eqn. for Schwarzschild space time and relate it to SUSY QM. Recently, it has been shown that FP eqn. corresponding to  Schwarzschild matric converge to the Schrodinger equation with harmonic oscillator potential. We applied the potential and power of isospectral deformation in SUSY QM to find a new version of the potential having identical eigenspectra with different eigen functions.

 \section{Fokker-Planck's Equation and Isospectral Deformation: Supersymmentric Connection}
  In the present section, we discuss about the dynamics of FP equation within the backdrop of SUSY QM. The FP equation for the Schwarzschild spacetime reduces to the Schrodinger equation as 
    \begin{equation}
     \pi T \Big[-\frac{\partial^2}{\partial \zeta^2}+\big(1+\zeta^2\big) \Big]\Psi(x)= E\, \Psi(x),
\end{equation}
where, effective potential $V = 1+ \zeta^2$ is the scaled  oscillator potential with, $\zeta= \sqrt{\pi T/D}\,y$ and $y=x- \Big(1/4\pi T\Big)$. Now, we exploit the attributes of isospectral deformation in SUSY QM to generate a new potential \cite{ara2, roy}. The superpotential corresponding to the potential in eqn. (17): $W(\zeta) = \zeta$. As discussed in the previous Section, we translate the superpotential by a general function as 
\begin{equation}
   W(\zeta) \longrightarrow \tilde W(\zeta) = W(\zeta) + f(\zeta),
\end{equation}
the above equation, under the Bernouli equation yields
\begin{equation}
    f(x)=\frac{2e^{-\zeta^2}}{\sqrt{\pi} (erf(\zeta)+\lambda)}.
\end{equation}
Using Eqn. (18) and (19), one gets the new superpotential 
\begin{eqnarray}
    \tilde W(x) = \zeta + \frac{2e^{-\zeta^2}}{\sqrt{\pi} (erf(\zeta)+\lambda)}.
\end{eqnarray}
The above superpotential has different functional form but it keeps the original potential $\tilde V_-=V_-$ intact. However, exploiting the eqns. (17), (18) and (20) one will get a deformed potential $\tilde V_+$, as following
\begin{equation}
 \hskip -0.7 cm \tilde V_+= 1+ \zeta^2+\frac{2e^{-\zeta^2}}{\sqrt{\pi}(erf (\zeta)+1+\lambda)} \Bigg[\frac{e^{-\zeta^2}}{\sqrt{\pi}(erf(\zeta)+1+\lambda)}+ 2\zeta\Bigg],
\end{equation}
with $\lambda$ as a constant parameter. It should be noted that the above potential will generate the same eigenvalues as original potential given in eqn. (7). Now, one can find the ground state ($\psi^{(0)}_1(\zeta)$) and first excited state ($\psi^{(1)}_1$) by doing simple mathematical calculation as:
\begin{eqnarray}
&& \tilde \psi^{(0)}_1(\zeta) \propto  \frac{2e^{-\zeta^2/2}}{erf(\zeta) + 1 +\lambda}, \qquad \quad \\
 &&\tilde \psi^{(1)}_1 \propto 2\zeta e^{\zeta^2/2} + \sqrt{\frac{1}{\pi}} \frac{2e^{-3\zeta^2/2}} {erf (\zeta) + 1 + \lambda}.
\end{eqnarray}
In a similar manner, using the creation operator,
\begin{equation}
    \tilde A^\dagger= \zeta- \sqrt{\frac{1}{\pi}}\frac{-2e^{\zeta^2}}{erf (\zeta) +1+\lambda}+ \frac{d}{d\zeta},
\end{equation}
one can obtain the $n^{th}$ excited state as,
\begin{equation}
  \hskip -0.7cm  \tilde \psi^{(n)}_1\propto \Big[\zeta+\sqrt{\frac{1}{\pi}} \frac{2e^{-\zeta^2}} {erf (\zeta) + 1 + \lambda}- \frac{d}{d\zeta}\Big] \nonumber\\
    \Big[\zeta+\frac{d}{d\zeta}\Big] \Big[-\zeta + \frac{d}{dx}\Big]^n e^{-\zeta^2/2}.
\end{equation}
The isospectral potential, together with its corresponding ground and first excited state wavefunctions, is shown for multiple values of the parameter $\lambda$
\begin{figure}[h]
\centering
\subfloat[]{\includegraphics[width = 4.3in]{Pot.png}} \vskip 0.8cm
\subfloat[]{\includegraphics[width = 3.2in]{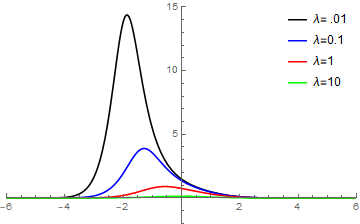}}
\subfloat[]{\includegraphics[width = 3.2in]{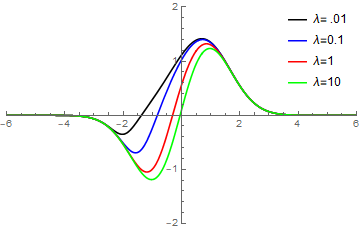}}
\caption{(a) Isospectral potential $\tilde V_+(x)$ (b) Ground state wavefunction for isospectral potential (c) First excited state for for various values of parameter $\lambda$.  }
\label{some example}
\end{figure}\\
As depicted in Figure, the isospectral potentials strongly depend on the parameter 
$\lambda$ near the origin. For small to moderate 
$\lambda$, the potentials differ noticeably, but these differences diminish at larger 
$\lambda$, converging to a universal asymptotic form. Similarly, the associated wavefunctions vary significantly near $x=0$ in amplitude and curvature, yet all satisfy normalizability by vanishing at infinity. Notably, the magnitude and positions of wavefunction nodes shift with 
$\lambda$, highlighting its critical influence on local quantum state properties despite the invariance of the energy spectrum.

\section{Conclusions}
This paper highlights the significance of isospectral deformation within the framework of SUSY QM. Specifically, we investigate the FP equation associated with the Schwarzschild metric by employing the SUSY QM formalism, wherein the corresponding thermal potential effectively reduces to an oscillator-type potential. The wavefunctions corresponding to the isospectral Hamiltonian exhibit structural features that are distinctly different from those of the original Hamiltonian  $H$. For varying values of  parameter  $\lambda$, the isospectral potential displays pronounced deviations in the vicinity of the origin. This parameter $\lambda$ modulates the strength and character of the deformation, thereby directly affecting the local potential profile. Correspondingly, the wavefunctions demonstrate significant variation near the origin; however, despite these local differences, the wavefunctions associated with vanish asymptotically, preserving square integrability and ensuring their physical validity. As a result, the probability density and node distribution are highly sensitive to $\lambda$, especially in regions where the potential undergoes substantial modification. These findings elucidate the subtle interplay between spectral isospectrality and spatial non-equivalence, with potential applications in quantum control and spectral design where tailoring local wavefunction properties without altering the global spectrum is desired. It should be noted that the effective potential is obtained from inverted oscillator thermal potential of Schwarzschild spacetime within the framework of FP formalism. The inverted oscillator plays a significant role in describing the cosmological singularities and inflationary scenarios. The isospectral and $\mathcal{PT}-$symmetric behavior of inverted harmonic oscillator are under investigation and will be reported elsewhere.

\end{document}